\documentclass[aps,epsfig,showpacs]{revtex4}

\usepackage{graphicx}

\begin{document}

\title{Shell-model description of monopole shift in neutron-rich Cu}
\author{N.A. Smirnova, A. De Maesschalck,  A. Van Dyck, K. Heyde}
\address{Vakgroep Subatomaire en Stralingsfysica, Universiteit Gent,
Proeftuinstraat 86, B-9000 Gent, Belgium}

\date{\today}

\begin{abstract}

Variations in the nuclear mean-field, in neutron-rich nuclei, are
investigated within the framework of the nuclear shell model. The
change is identified to originate mainly from the monopole part of
the effective two-body proton-neutron interaction. Applications
for the low-lying states in odd-$A$ Cu nuclei are presented. We
compare the results using both schematic and realistic forces. We
also compare the monopole shifts with the results obtained from
large-scale shell-model calculations, using the same realistic
interaction, in order to study two-body correlations beyond the
proton mean-field variations.
\end{abstract}


\maketitle



The nuclear shell model is based on the assumption that each of
the nucleons in an atomic nucleus moves independently in an
average field created by the other nucleons. One of the major
successes of the model, even in its simplest version when the
average potential is approximated by the the harmonic oscillator
potential with the centrifugal term ($l\cdot l$) and spin-orbit
interaction ($l \cdot s$), was the reproduction of the so-called
magic numbers, i.e. the numbers of protons and neutrons which give
additional stability to a nuclear system (2,8,20,28,50,82,126).
Inclusion of correlations beyond the mean-field approximation
through the introduction of the residual two-body interaction
provides us with an excellent description of atomic nuclei in the
vicinity of the valley of $\beta $-stability~\cite{BrWi88,CaNo99}.

The shell structure of nuclei with large proton or neutron excess
differs considerably from that of the stable nuclei. The important
role in the mechanism of changes is played by the proton-neutron
interaction as has long been recognized~\cite{DeSh53,FePi79}. It
was demonstrated by the mean-field calculations~\cite{Dob94} that
the neutrons in very neutron-rich nuclei experience an average
field with a much more diffused boundary due to the close-lying
continuum and increased importance of the pairing interaction. As
a result, the shell structure resembles that of a harmonic
oscillator with the spin-orbit term only.

Within the shell-model concept, the description of the nuclear
mean field is conventionally obtained by considering the so-called
{\it monopole Hamiltonian} constructed from the centroids of the
two-body interaction~\cite{BaFr64,PoZu81,HeDeCo94,DuZu96}. The
eigenvalues of this Hamiltonian were shown~\cite{DuZu96} to
provide the average energies of the specific spherical
configurations. They are usually referred to as {\it effective
single-particle energies}, and should reproduce the energies of
single-proton (single-neutron) states in odd-$A$ nuclei with $Z$
($N$) equal to a magic number plus or minus one proton (neutron).

Recent shell-model calculations of neutron-rich nuclei with masses
below $A\approx 40$ studied the mechanisms of the changes in the
shell-structure of the neutron mean-field~\cite{OtFu01,OtUt02}. It
was shown that an increasing $N/Z$ ratio, certain neutron orbitals
move up or down thus deteriorating the usual shell closures and
creating new ones. For example, the well-pronounced  shell gap at
$N=20$ in nuclei near $^{40}$Ca, disappears for nuclei with
$Z<12$, while $N=16$ instead appears as a new magic number. This
is a consequence of the crucial role played by the attractive
proton-neutron effective nucleon-nucleon interaction (especially,
in $T=0$ channel) in which spin-orbit partners are
involved~\cite{FePi79}.

It can be expected that the proton-neutron interaction will
influence in a major way the effective single-proton energies of
the proton mean-field. It is well-known that the ordering and
spacing of the proton mean-field orbitals in neutron-deficient or
neutron-rich isotopes is not the same compared to $\beta $-stable
nuclei, and may exhibit pronounced differences.

In this article, we intend to investigate the evolution of the proton
mean-field from nuclei, situated around the valley of $\beta
$-stability, towards very neutron-rich systems. In particular, we
consider nuclei near semi-magic ones, namely, those which differ
from a semi-magic system by an extra proton particle (hole).
Typical examples are the F, Sc, Cu, Sb and Bi isotopes.

The advantage of this choice is that the low-lying states in these
nuclei usually carry a major part of the proton single-particle
strength and thus a direct comparison between theory and
experiment becomes possible. These nuclei can be rather well
described by an inert core with one extra proton particle (hole)
and a number of valence neutrons outside the core. We then
estimate the effective single-particle energies ($\tilde
\epsilon_{j}$) in the proton mean-field as a bare single-particle
energy ($\epsilon_{j}$) plus the average proton-neutron part of
the interaction~\cite{Goo77,Sor84,HeJo87}. In this approximation,
the evolution of the single-particle states as a function of
neutron number will depend on the specific proton-neutron
effective interactions 
($\langle j_\pi j_\nu |V| j_\pi j_\nu \rangle $) 
and the pairing properties amongst the neutrons.

The existence of the single-proton (single-neutron) shifts
is well known experimentally in a series of isotopes (isotones)~\cite{Franchoo}.
Recently a lot of new data has been obtained on
odd-$A$ F~\cite{Sak99,ThBa03,PeBa03}, Cu~\cite{Fra01} and Bi~\cite{KuPl03,Hilde}
isotopes, up to very neutron-rich species.

In the present study we apply this approach to neutron-rich Cu-isotopes.
We choose two types of the effective interaction, a
schematic zero-range force and a realistic one, based on the
$G$-matrix~\cite{MHJ95,Frederic}, in order to test the sensitivity
to the use of various (schematic versus realistic) forces.
Finally, a comparison of the monopole shifts with the results of
the large-scale shell-model diagonalization is carried out.\\[5mm]


The interest in neutron-rich nuclei around $^{68}$Ni and in
particular the questioned magicity of $^{78}$Ni has recently
motivated the global study 
of low-energy nuclear structure for both odd-$A$ Cu nuclei~\cite{Fra01} 
at the LISOL facility in Louvain-la-Neuve and 
odd-odd mass Cu at ISOLDE, Cern~\cite{JVR} up to very neutron-rich species.
Figure~1 contains the
selected states of odd-$A$ Cu with spin and parity quantum numbers
of the valence proton oscillator shell above $Z=28$: $2p_{3/2}$,
$1f_{5/2}$, $2p_{1/2}$, $1g_{9/2}$. The ground state of odd-$A$
Cu-nuclei up to $^{73}$Cu is $3/2^-$ conform with the occupation
of the $2p_{3/2}$ oscillator orbital. As seen from Fig.~1, the
$5/2^-$ state starts to drop drastically in energy for nuclei with
$A>69$ and might be expected to become the ground state in heavy
odd-$A$ Cu.

Since we are interested in the variation of the proton
single-particle energies, it is consistent to compare our
theoretical results with the experimental centroids of the
$2p_{3/2}$, $1f_{5/2}$, $2p_{1/2}$ and $1g_{9/2}$ single-proton
states, defined as
\begin{equation}
\label{energy_w}
E (nlj) = \frac{\sum_{k}S^k_{nlj} E^k_{nlj}}{\sum_{k}S^k_{nlj}} ,
\end{equation}
where $S^k_{nlj}$ are spectroscopic factors for a proton transfer
(from $(nlj)$ single-particle state), and $E^k_{nlj}$ is the
experimental excitation energy of the $k$th state with $J=j$ (both
taken from \cite{nndc,nds59,nds61,nds63,nds65,ZeNo78} and Ref.~\cite{Fra01}).

Unfortunately, the data on spectroscopic factors is rather 
scare and differs significantly from one experiment to another. 
In Table I we summarize the available spectroscopic strength
for the states of interest in $^{59 - 65}$Cu (see captions of the tables for references).
The experimental centroids are plotted in Fig.~2(a). For those states
(isotopes) for which no experimental data exists, only the lowest states of a
given spin is indicated.
As seen from the table, the sum rule (for the transfer to both,
$T_f=T_i \pm \frac12$ states) is far from being accomplished,
especially, for $9/2^+$ states. Thus, one should be cautious with
the trends of experimental ``centroids'' for this particular state.\\[5mm]


The low-lying states of a nucleus with one proton (neutron)
outside the doubly magic core can be considered as mainly
single-particle states and they usually serve to obtain
information on the single-particle energies $\epsilon _j$ 
used in the shell-model. Here $j$ stands for the quantum
numbers $(n,l,j,m)$ of a given orbital.

The interaction between the valence proton and the filled neutron
orbital results in a change of the single-particle energy that can
be expressed as~\cite{DeShalit,FePi79}
\begin{equation}
\tilde{\epsilon}_{j_{\pi }} = \epsilon_{j_{\pi }}+
n_{\nu } E_{j_{\pi } j_{\nu }},
\end{equation}
where $n_{\nu }$ is the number of valence neutrons and $E_{j_{\pi
}j_{\nu }}$ is the average matrix element of the proton-neutron
interaction:
\begin{equation}
\label{centroid}
E_{j_{\pi }j_{\nu }} =
\frac{\sum_J \langle j_{\pi } j_{\nu }; J |V| j_{\pi }j_{\nu }; J\rangle
(2J+1)}{\sum_J (2J+1)} \, ,
\end{equation}
with $V$ standing for the effective two-body interaction, and $J$
being the total angular momentum of a two-body configuration.

Taking into account that the model space for neutrons in general
consists of several single-particle orbitals and that the pairing
force spreads neutrons over these valence orbitals, one arrives at
the following expression~\cite{Goo77,Sor84,HeJo87}:
\begin{equation}
\label{espe}
\tilde{\epsilon}_{j_{\pi }} = \epsilon_{j_{\pi }}+ \sum_{j_{\nu }}
E_{j_{\pi }j_{\nu }} (2 j_{\nu } + 1) v^{2}_{j_{\nu }}\; ,
\end{equation}
where $v^{2}_{j_{\nu }}$ stands for the occupation probability of
the orbital $j_{\nu }$ and the summation is performed over all
valence neutron orbitals.

From eq.~(\ref{espe}) it is obvious that the proton-neutron
interaction causes the proton single-particle states to change,
depending on its character (attraction or repulsion) and
magnitude. The larger the overlap between the proton and neutron
radial wave functions and the higher $j_{\nu }$ of a given neutron
orbital is, the stronger the variations are.

Remark that only the monopole component from the multipole
representation of the effective force contributes
to the average matrix element (\ref{centroid})~\cite{HeJo87}.
This is why the effect is referred to as the {\it monopole shift}.

The basic ingredients required in (\ref{espe}) are:
(i) the shell-model space; (ii) the effective proton-neutron interaction, and
(iii) neutron occupation probabilities.

For the description of the Cu-nuclei we have chosen the {\it
shell-model space} consisting of four valence orbitals,
$2p_{3/2}$, $1f_{5/2}$, $2p_{1/2}$, and  $1g_{9/2}$, both for
neutrons and protons, outside $^{56}$Ni-core. This model space
contains all necessary degrees of freedom to describe low energy
nuclear structure.

In the present work we have restricted ourselves
to two types of the {\it effective interaction}.
First, we consider the zero-range $\delta $-interaction
with the spin-exchange term ~\cite{Heyde94}
\begin{equation}
\label{delta}
 V = -V_0 \delta (\vec{r}_{\pi } -\vec{r}_{\nu }) (1
-\alpha + \alpha \vec{\sigma }_{\pi }\vec{\sigma }_{\nu }) \,.
\end{equation}
The two parameters, the overall strength $V_0$ and the strength of
the exchange term $\alpha $ are to be adjusted to the experimental
data in odd-odd nuclei from the mass region one is studying.

The geometrical properties of the zero-range interaction, even
including spin-exchange contributions  (\ref{delta}), allow to obtain a
very simple analytical expression for the monopole shift:
\begin{equation}
\label{espe_delta}
\tilde{\epsilon}_{j_{\pi }} = \epsilon_{j_{\pi }}-V_0(1-\alpha )
\sum_{j_{\nu }}F^0_{n_{\pi }l_{\pi };n_{\nu }l_{\nu }} (2j_{\nu } +1)v^{2}_{j_{\nu}}\; .
\end{equation}
where $F^0_{n_{\pi }l_{\pi };n_{\nu }l_{\nu }}$ are Slater
integrals. In order to derive this result, we had to make use of a
peculiar property of the Wigner $3j$-symbols, as described in
Appendix A.

We stress the fact that since the average proton-neutron matrix element from
 the $\delta $-interaction entering (\ref{espe_delta}) does not depend on $j_{\pi }$,
the relative shift for spin-orbit partners 
(see e.g. the result for $2p_{1/2}$ and $2p_{3/2}$ orbitals in Fig.~2(b)) 
stays constant and remains equal
to the difference of the original single-particle energies
$(\epsilon_{j_{\pi }}-\epsilon_{j'_{\pi }})$.

For the Cu-isotopes, we have determined the parameters of
(\ref{delta}) to be $V_0 = 400$ MeV.fm$^3$ and $\alpha = 0.1$.
These values suit the systematics known from the description of
heavier isotopes~\cite{HeJo87} and they result in two-body matrix
elements of the same order as those of the realistic interaction
which will be described below. The chosen interaction also gives a
reasonable, although slightly stretched, spectrum of
$^{58}$Cu~\cite{LiPi03}.

The second type of interaction used here is a realistic one,
obtained from the $G$-matrix through a Brueckner-Hartree-Fock
procedure~\cite{MHJ95} and modified further for a monopole
correction~\cite{Frederic} following the recipe of
Ref.~\cite{PoZu81} to account for the saturation properties. The
interaction works very well in the description of Ni and
Cu-isotopes, reproducing many known spectroscopic
properties~\cite{Frederic}.

The neutron {\it occupation probabilities} $ v^{2}_{j_{\nu }}$
have been obtained from the solution of the BCS equations using a
pairing Hamiltonian~\cite{Heyde94}:
\begin{equation}
\label{pairing}
 H = -G\sum_{jm,j'm'} a^{\dagger }_{jm}a^{\dagger }_{j,-m}
 \tilde a_{j'm'} \tilde a_{j',-m'} \; .
\end{equation}
The strength of the pairing interaction was chosen to reproduce
the pairing gaps in neighboring even-even Ni-isotopes~\cite{OrMa00} resulting
in a dependence $G=23/A$ MeV.\\[5mm]


In Fig.~2 we compare the experimental energy centroids (Fig.~2(a)) with the results
of the calculations according to eq.~(\ref{espe}) with
two types of the interaction (Fig.~2(b,c)).

Figure~2(b) contains the results obtained using the $\delta
$-force. For $^{57}$Cu, the lowest states are given by the
empirical values $\epsilon_{j_{\pi }}$. These values are modified
by the two-body interaction. Effective single-proton energies for
the $1f_{5/2}$ and $1g_{9/2}$ states first move up, when neutrons
are filling the $pf$ orbitals, and then (from $^{69}$Cu) go down
relative to the $2p_{3/2}$ state as the neutron $1g_{9/2}$ orbital
starts to be filled. This is governed by the relative magnitude of
the centroids of the interaction. In particular, the drastic
lowering of the $1f_{5/2}$ and $1g_{9/2}$ orbitals stems from more
attractive $E_{1f_{5/2} 1g_{9/2}}$ and $E_{1g_{9/2} 1g_{9/2}}$
values relative to $E_{2p_{3/2} 1g_{9/2}}$. The energy of the
$2p_{1/2}$ orbital stays unchanged with respect to the $2p_{3/2}$
ground state as has been mentioned before. At $A=79$ we observe a
crossing between the effective $2p_{1/2}$ and $1f_{5/2}$ states,
however, the chosen interaction does not reproduce the expected
crossing of the effective $2p_{3/2}$ and $1f_{5/2}$ orbitals.

In Fig.~2(c), we present the results obtained with a realistic
effective interaction described above. Only diagonal
proton-neutron matrix elements contribute to the expression
(\ref{espe}). Though the overall trend obtained from both types of
the interaction remains the same, as follows from the  similar
character of the matrix elements, there are certain distinctions.
First, the energy of the $2p_{1/2}$ orbital varies relative to the
$2p_{3/2}$ ground state orbital (see Fig.~2(c)). Secondly, the
proton $1f_{5/2}$ orbital decreases steeper as neutrons are added
in the $1g_{9/2}$ orbital and for $^{79}$Cu the inversion of the
$1f_{5/2}$ and $2p_{3/2}$ orbitals is predicted, compared to the
use of the $\delta $-interaction.

We point out that the lack of agreement between the variation of
the effective $1g_{9/2}$ single-proton state and the experimental
$9/2^+$ level is due to the fact that the single-particle
strength of the $1g_{9/2}$ state lies higher up and the lowest
state which is plotted in Fig.~2(a), due to the lack of
experimental data, contains only a small fraction of that 
(see Table I).

In order to get an insight into additional correlations that cause a
variation of the single-particle centroids
of $2p_{3/2}$, $1f_{5/2}$, $2p_{1/2}$ and $1g_{9/2}$ proton states
beyond the monopole shift, we have
carried out large-scale shell-model calculations using the same
realistic effective interaction as described above. The model
space consists of the $(1f_{5/2}2p_{3/2}2p_{1/2}1g_{9/2})$
orbitals outside the $^{56}$Ni-core, truncated in case of $^{67}$Cu,
$^{65,69}$Cu and $^{71}$Cu (up to 3, 4 and 8 particles, respectively, 
were allowed to occupy the $1g_{9/2}$ orbital). 
The calculations were done using the shell-model code Oxbash~\cite{Oxbash}.
The single-particle centroids for the $2p_{3/2}$,
$1f_{5/2}$, $2p_{1/2}$ and $1g_{9/2}$ states obtained by taking into account
up to about 400-800 excited states weighted by their spectroscopic factors 
(in order to exhaust the sum rule for $T_f=T_i -\frac12$ transfer up to
98\% or to get a reasonable saturation of the total value for truncated
spaces) are shown in Fig.~2(d). The wave functions of the corresponding
Ni-isotopes have been calculated without any restrictions of the model space.

A remarkable result is that the centroids shown in Fig.~2(d) are rather close 
to those obtained
by the analytical approach (\ref{espe}) as shown in Fig.~2(c)
(the wiggles in the behavior of the shell-model $1g_{9/2}$-centroid in Fig.~2(d)
for $^{65-69}$Cu are related most probably to the truncations of the model
space which is otherwise too large to be treated numerically).
The monopole energy shifts incorporates the effects of a certain class of residual 
interactions (self-energy correction of the proton single-particle energy because 
of proton-neutron interactions and neutron pairing) on the bare proton 
single-particle energy  within the model space, as expressed by eq.~(\ref{espe}).  
 
A comparison of these results thus illuminates the question what type of correlations 
is important in the calculation of the energy shifts beyond the monopole
shift. A rather good agreement between the centroids of Figs.~2(c) and 2(d)
confirms that the basic ingredients of the single-particle centroid variations
are already taken into account by eq.~(\ref{espe}).
This is a most interesting and valuable result.

To illustrate the importance of the pairing effects, we plot in Fig.~2(e) the
effective single-proton energies obtained from the monopole Hamiltonian
with normal filling of the single-particle orbitals by neutrons (the approach
chosen in Refs.~\cite{BaFr64,PoZu81,DuZu96,OtFu01,OtUt02}).
It is seen that through the occupation probabilities,
the pairing force (Fig.~2(c))
smoothes significantly the single-particle behavior.
Fig.~2(f) shows the same as Fig.~2(e), but with a different order of filling
of neutrons orbitals (neutron $2p_{1/2}$ is filled before neutron $1f_{5/2}$).
The behavior of single-particle orbitals does not have a zigzag structure 
in this case, but the maximum of the centroids is shifted to $^{63}$Cu,
compared to the calculations with pairing effects (Fig.~2(b,c)).
Let us note that physically the order of filling chosen in Fig.~2(e) is more
realistic than that implied in Fig.~2(f), 
as follows from the experimental spectra of $^{67,69}$Ni
(neutron single-particle orbital $2p_{1/2}$ is higher than $1f_{5/2}$ in these
nuclei).

Since the lowest-lying $J^{\pi }$ = $1/2^-$, $3/2^-$,
$5/2^-$ and $9/2^+$ states represent only a fraction of the total
proton single-particle strength, we compare
the energies of the lowest $1/2^-$, $3/2^-$, $5/2^-$ 
and $9/2^+$ experimental states (Fig.~3(a)) and theoretical states obtained
from the large-scale diagonalization (Fig.~3(b)). In
general, we remark a very good agreement with the experimental
data due to the quality of the effective interaction.

By comparing the results of Figs.~2(c,d) and 3(b), it is clear that
the correlations beyond the monopole approximation are necessary
to understand a detailed fine structure of the energy variations for the
lowest $1/2^-$, $5/2^-$ and $9/2^+$ levels. This is particularly
so for the $1/2^-$ and $5/2^-$ levels where specific kinks, related
to a possible shell closure at $N=40$, are clearly seen in the
diagonalization. The general trend of the $9/2^+$ level, as
obtained from diagonalization, is quite different compared to the
monopole shift. These effects are obviously a result of
interactions of the pure single-particle configurations with many
of the excited configurations with the same $J^{\pi }$ value.

Finally, in Fig.~4 we summarize the spectroscopic factors of
the $2p_{3/2}$, $1f_{5/2}$,  $2p_{1/2}$ and $1g_{9/2}$ proton stripping
from the $^{A-1}$Ni-core into the lowest-lying $1/2^-$, $3/2^-$, $5/2^-$ and $9/2^+$
states of $^A$Cu, respectively, as provided by the shell-model diagonalization. 
The spectroscopic factors have been normalized to the sum rule presented in
Appendix B. The absolute spectroscopic factors for the lowest states of a
given spin are shown in Table I in comparison with the available experimental
data on $^{59-65}$Cu. The calculated spectroscopic factors are in rather good
agreement for the lowest $3/2^-$ and $1/2^-$ states (within 20\%), however,
the disagreement increases for $5/2^-$ and $9/2^+$ states in these isotopes.
It is not clear, at present, what the origin of these differences is. It may
be due to a lack of experimental data, but it may also point out certain
deficiencies in the realistic interaction presently used.
As for the whole series of Cu-isotopes, the calculations show 
that only the $3/2^-$ ground state remains more
or less a pure single-particle state. The lowest $1/2^-$ state is
predicted to carry about half of the $2p_{1/2}$ single-particle
strength. The lowest $5/2^-$ state shows much more mixing for
$^{61-69}$Cu. This is in line with experimental observations of a
few $5/2^-$ states in these isotopes even at low energies.
Finally, the $9/2^+$ state is predicted to be a rather pure
single-particle $1g_{9/2}$ state only for light isotopes. Starting
from $^{67}$Cu, the single-proton strength almost vanishes. In
order to test these predictions, more precise data on
spectroscopic factors and level spin and parity assignments would
be very helpful.


In conclusion, it becomes clear  that using both simple effective and 
realistic forces in order to derive the corresponding monopole energy shifts, 
the experimental proton single-particle centroids are, at present, 
only moderately described. 
This comparison is hampered by the lack of reliable and extensive experimental results on 
one-nucleon transfer reactions. 
 
An interesting point is a reproduction of a lowering in the $1f_{5/2}$ energy
centroid, once the $N=40$ neutron number is passed, in particular using
realistic forces. 
We also remark that correlations that define the monopole energy shift, 
as defined in eq.~(\ref{espe}), are very much consistent 
with the full theoretical single-particle energy centroid resulting from a 
diagonalization of the full energy matrix, encompassing all possible residual interactions 
within the large model space. This result is an interesting one and has to be studied 
in other mass regions, too. 
Although the energies obtained from  the diagonalization 
in the large model space concerning the lowest state of
each spin, parity, following rather well the corresponding experimental energies
and exhibit interesting fine structure, the theoretical centroid, using 
a significant number of the eigenstates, becomes rather smooth again.
At the same time, more experimental data on structure of nuclei from 
the region would serve a good testing ground for the present
results and the interactions used.

\acknowledgements{
N.A.S. thanks E.~Caurier and F.~Nowacki from IReS (Strasbourg)
for making available the interaction exploited here,
and for useful discussions. We thank R.F.~Casten for interesting discussions.
This work was supported by the Inter-University Attraction Poles (IUAP) under
project P5/07. K.H. is grateful to the FWO-Vlaanderen for financial
support.} \\[5mm]

\begin{center}
{\bf Appendix A: Properties of the  $\delta $-interaction}
\end{center}

The diagonal matrix element of the interaction (\ref{delta})
between the two-body states in $jj$-coupling reads
\begin{equation}
\langle j_{\pi } j_{\nu }; J|V|j_{\pi } j_{\nu }; J \rangle =
-V_0 (2j_{\pi } + 1)(2j_{\nu } + 1) ((1 -2\alpha -2\alpha (-1)^J)
\left(
\begin{array}{ccc}
j_{\pi } & j_{\nu } & J \\
1/2 & -1/2 & 0 \\
\end{array}
\right)^2
+\left(
\begin{array}{ccc}
j_{\pi } & j_{\nu } & J \\
1/2 & 1/2 & -1 \\
\end{array}
\right)^2 ) \, .
\end{equation}

In order to evaluate the average matrix element,
\begin{equation}
E_{j_{\pi }j_{\nu }} =
\frac{\sum_J \langle j_{\pi } j_{\nu }; J |V| j_{\pi }j_{\nu }; J\rangle
(2J+1)}{\sum_J (2J+1)} \, ,
\end{equation}
using the above two-body proton-neutron matrix elements,
we have to exploit the well-known summation properties for $3j$-symbols,
\begin{equation}
\sum_{J} (2J+1)
\left(
\begin{array}{ccc}
j_{\pi } & j_{\nu } & J \\
1/2 & -1/2 & 0 \\
\end{array}
\right)^2  = 1 \, ,
\end{equation}
\begin{equation}
\sum_{J}  (2J+1)
\left(
\begin{array}{ccc}
j_{\pi } & j_{\nu } & J \\
1/2 & 1/2 & -1 \\
\end{array}
\right)^2  = 1 \, .
\end{equation}
Besides, in the calculations the particular expression,
\begin{equation}
\label{phase}
\sum_{J} (-1)^J (2J+1)
\left(
\begin{array}{ccc}
j_{\pi } & j_{\nu } & J \\
1/2 & -1/2 & 0 \\
\end{array}
\right)^2  \, ,
\end{equation}
appears for which we have found no standard result in the
literature on angular momentum algebra (e.g., Ref.~\cite{VaMo88}). 
We had to be able to establish that the
above expression (\ref{phase}) vanishes, as in the most general
case,
\begin{equation}
\sum_{J} (-1)^J (2J+1)
\left(
\begin{array}{ccc}
j & j' & J \\ m & -m & 0 \\
\end{array}
\right)^2 =0  \, ,
\end{equation}
due to the peculiar summation properties of the considered $3j$-symbols
which we would like to mention here:
\begin{equation}
\sum_{J=even} (-1)^J (2J+1) \left(
\begin{array}{ccc}
j & j' & J \\ m & -m & 0 \\
\end{array}
\right)^2  = 1/2 \, ,
\end{equation}
\begin{equation}
\sum_{J=odd} (-1)^J (2J+1) \left(
\begin{array}{ccc}
j & j' & J \\ m & -m & 0 \\
\end{array}
\right)^2  = -1/2 \, .
\end{equation}
Such summations on the odd or even values of the angular momentum are often
encountered in quantum mechanics.

As a result, we obtain for the average matrix element of the proton-neutron interaction
the following analytical expression:
\begin{equation}
E_{j_{\pi }j_{\nu }} = -V_0 F^0_{n_{\pi }l_{\pi };n_{\nu }l_{\nu }}
(1- \alpha )\; ,
\end{equation}
where $F^0_{n_{\pi }l_{\pi };n_{\nu }l_{\nu }}$ stands for the Slater integral.

Remark, that the average matrix element does not depend on the $j$
quantum number of the orbitals involved, as well as it vanishes
for the spin-exchange term.

In a similar way, we can evaluate the average particle-hole matrix
element. The particle-hole matrix element relates to the
particle-particle one via the Pandya
transformation~\cite{Heyde94}:
\begin{equation}
\langle j_{\pi } j^{-1}_{\nu }; J|V|j_{\pi } j^{-1}_{\nu }; J \rangle =
-\sum_{J'} (2J' +1)
\left\{
\begin{array}{ccc}
j_{\pi } & j_{\nu } & J' \\
j_{\pi } & j_{\nu } & J  \\
\end{array}
\right\}
\langle j_{\pi } j_{\nu }; J|V|j_{\pi } j_{\nu }; J \rangle \; .
\end{equation}
Exploiting the following summation properties for $3j$-symbols,
\begin{equation}
\sum_{J'} (2J'+1)
\left\{
\begin{array}{ccc}
j_{\pi } & j_{\nu } & J' \\
j_{\pi } & j_{\nu } & J  \\
\end{array}
\right\}
\left(
\begin{array}{ccc}
j_{\pi } & j_{\nu } & J \\
1/2 & -1/2 & 0 \\
\end{array}
\right)^2  =
\left(
\begin{array}{ccc}
j_{\pi } & j_{\nu } & J \\
1/2 & 1/2 & -1 \\
\end{array}
\right)^2 \, ,
\end{equation}
\begin{equation}
\sum_{J'} (2J'+1)
\left\{
\begin{array}{ccc}
j_{\pi } & j_{\nu } & J' \\
j_{\pi } & j_{\nu } & J  \\
\end{array}
\right\}
\left(
\begin{array}{ccc}
j_{\pi } & j_{\nu } & J \\
1/2 & 1/2 & -1 \\
\end{array}
\right)^2  =
\left(
\begin{array}{ccc}
j_{\pi } & j_{\nu } & J \\
1/2 & -1/2 & 0 \\
\end{array}
\right)^2  \, ,
\end{equation}
and making use of the peculiar relation (\ref{phase}), one obtains
for the average particle-hole matrix element the result
\begin{equation}
E_{j_{\pi }j^{-1}_{\nu }} = V_0 F^0_{n_{\pi }l_{\pi };n_{\nu }l_{\nu }}
(1- \alpha )\; .
\end{equation}
This expression differs from that for a particle-particle average matrix
element only by its sign.

\begin{center}
{\bf Appendix B: Sum rule for the spectroscopic factors}
\end{center}

The total spectroscopic strength in proton stripping reactions
to the states with $T_f = T_i -\frac12$ 
is given by the following sum rule~\cite{BrGl77}:
\begin{equation}
\label{sum_rule}
C^2 \sum_f  \frac{2J_f+1}{2J_i+1} S_f(nlj) = 
\frac{T_i +T_{iz}}{2T_{iz} (2T_i + 1)} 
\left[(T_i +T_{iz}+1)\langle \tilde n_{\pi }(j) \rangle_{J_i,T_i}-
(T_i -T_{iz}+1)\langle \tilde n_{\nu }(j) \rangle_{J_i,T_i}\right] \, ,
\end{equation}
$J_i,J_f$ are the spin and $T_i,T_f$ are the isospin quantum numbers 
of the initial and final states (in $^{A-1}$Ni and $^A$Cu, respectively), 
$C = (T_i, T_{iz}, \frac12, -\frac12 |T_f=T_i-\frac12, T_{fz}=T_{iz}-\frac12)$ 
is the Clebsch-Gordan coefficient,
$\langle \tilde n(j) \rangle $ is the average number of proton or neutron holes 
in the $j$-orbital in the initial state (ground state of $^{A-1}$Ni in our
case).

Since in the case considered here, $J_i =0$, $J_f= j$, 
the sum rule (\ref{sum_rule}) reduces to
\begin{equation}
C^2 \sum_f S_f(nlj) (2J_f+1) = \langle \tilde n_{\pi }(j) \rangle_{T_i}-
\frac{1}{2T_i +1}\langle \tilde n_{\nu }(j) \rangle_{T_i} \, ,
\end{equation}
as indicated in Table I where the average number of neutron holes in
the ground state of Ni-isotopes are taken from the shell-model diagonalization.


\begin{table}
\begin{center}
\caption{Available experimental data on spectroscopic factors of
the lowest $3/2^-$, $5/2^-$, $1/2^-$ and $9/2^+$ states in $^{59,61,63,65}$Cu
(taken from taken from Ni(d,n) or Ni($^3$He,d) proton-stripping reaction 
studies~\protect\cite{nndc,nds59,nds61,nds63,nds65,ZeNo78}). The sum rule refers
to the proton transfer to both, $T_f = T_i \pm \frac12$ states, 
while the shell-model sum rule is given for the states with 
$T_f = T_i - \frac12$ only. 
The calculated spectroscopic factors for the
lowest states of a given spin are indicated as well.}
\begin{tabular}{cccccccccc}
& $I^{\pi }$ & ~~$E_{\rm exp}$ (MeV)~~ & ~$C^2 S(2J+1)$~ & ~$C^2 S(2J+1)$~
& ~~Sum rule~~& ~$C^2 S(2J+1)$~& ~~Sum rule~~ & ~~Centroid~~ & ~~Centroid \\
& & & (experiment) & total & & (shell model) & (shell model) 
& absolute & relative \\
& & & & (experiment) & &  &  & (experiment) & (experiment) \\
\hline
$^{59}$Cu~~ & $\frac{3}{2}^-$ & 0 & 1.85 & & & 2.07 & & & \\ 
             &   & 3.130 & 0.22 & & & & & & \\
             &   & 3.742 & 0.11 & & & & & & \\
             &   & 3.886 & 0.51 & 2.39 & 4 & & 2.974 & 1.146 & 0 \\ 
\hline
& $\frac{5}{2}^-$ & 0.914 & 2.5 & & & 2.43 & & & \\ 
             &   & 4.307 & 1.76 & 4.26 & 6 & & 4.252 & 2.316 & 1.170 \\ 
\hline
& $\frac{1}{2}^-$ & 0.491 & 0.84 & & & 0.67 & & & \\ 
             &     & 4.000 & 0.30 & & & & & & \\
             &     & 4.349 & 0.35 & & & & & & \\
             &   & 5.231 & 0.21 & 1.70 & 2 &  & 1.384 & 2.490 & 1.345 \\ 
\hline
& $\frac{9}{2}^+$ & 3.043 & 2.4 & 2.4 & 10 & 4.31 & 6.724 & 3.043 & 1.897 \\ 
\hline
$^{61}$Cu & $\frac{3}{2}^-$ & 0 & 2.16 & & & 2.35 & & & \\ 
          &      & 1.933 & 0.22 & 2.38 & 4 &  & 3.539 & 0.179 & 0 \\ 
\hline
& $\frac{5}{2}^-$ & 0.970 & 3.26 & & & 1.38 & & & \\ 
             &   & 1.394 & 0.36 & & & 0.62 & & & \\
             &   & 1.904 & 0.06 & & & & & & \\
             &   & 2.203 & 0.56 & & & & & & \\
             &   & 2.793 & 0.13 & 4.37 & 6 &  & 5.125 & 1.230 & 1.051 \\ 
\hline
& $\frac{1}{2}^-$ & 0.475 & 0.96 & & & 0.8 & & & \\ 
             &   & 2.089 & 0.06 & 1.02 & 2 & 0.008 & 1.668 & 0.570 & 0.391 \\ 
\hline
& $\frac{9}{2}^+$ & 2.721 & 2.80  & & & 4.04 & & & \\
             &    & 3.276 & 0.14  & 2.94 & 10 & & 8.069 & 2.747 & 2.569 \\ 
\hline
$^{63}$Cu & $\frac{3}{2}^-$ & 0     & 3.0 & & & 2.62 & & & \\ 
          &       & 1.547 & $<0.01$ & & & & & & \\
          &       & 2.012 &  0.12 & & & & & & \\
          &       & 2.780 &  0.16 & 3.28 & 4 & & 3.778  & 0.209 & 0 \\ 
\hline
& $\frac{5}{2}^-$ & 0.962 & 1.92 & & & 0.40 & & & \\ 
           &     & 1.412 & 2.70 & & & 1.56 & & & \\
           &     & 2.337 & 0.48 & & & 0.78 & & & \\
           &     & 3.225 & 0.30 & 5.4 & 6 & & 5.489 & 1.226 & 1.435 \\ 
\hline
& $\frac{1}{2}^-$ & 0.670 & 1.38 & & & 0.90 & & & \\ 
           &     & 2.062 & 0.29 & 1.67 & 2 & & 1.800 & 0.912 & 0.703 \\ 
\hline
& $\frac{9}{2}^+$ & 2.506 & 5.30  & & & 3.24 & & & \\
           &      & 3.970 & 0.51 & 5.81 & 10 & & 8.648 & 2.635 & 2.425 \\ 
\hline
$^{65}$Cu &  $\frac{3}{2}^-$ & 0 & 3.08 & & & 2.92 & & & \\ 
          &      & 1.725 & 0.06 & & & 0.07 & & & \\
          &      & 2.329 & 0.24 & 3.38 & 4 & & 3.898 & 0.196 & 0 \\ 
\hline
& $\frac{5}{2}^-$ & 1.113 & 1.14 & & & 0.21 & & & \\ 
           &     & 1.623 & 2.40 & & & 2.13 & & & \\
           &     & 2.107 & 0.36 & 3.9 & 6 & & 5.694 & 1.519 & 1.323 \\ 
\hline
& $\frac{1}{2}^-$ & 0.770 & 1.30 & & & 1.01 & & & \\ 
           &     & 2.213 & 0.34 & 1.64 & 2 & 0.15 & 1.883 & 1.069 & 0.873 \\ 
\hline
& $\frac{9}{2}^+$ & 2.526 & 3.80 & 3.80 & 10 & 0.76 & 8.970 & 2.526 & 2.330 \\ 
\hline
\end{tabular}
\end{center}
\end{table}


\begin{figure}
\includegraphics[width = \textwidth]{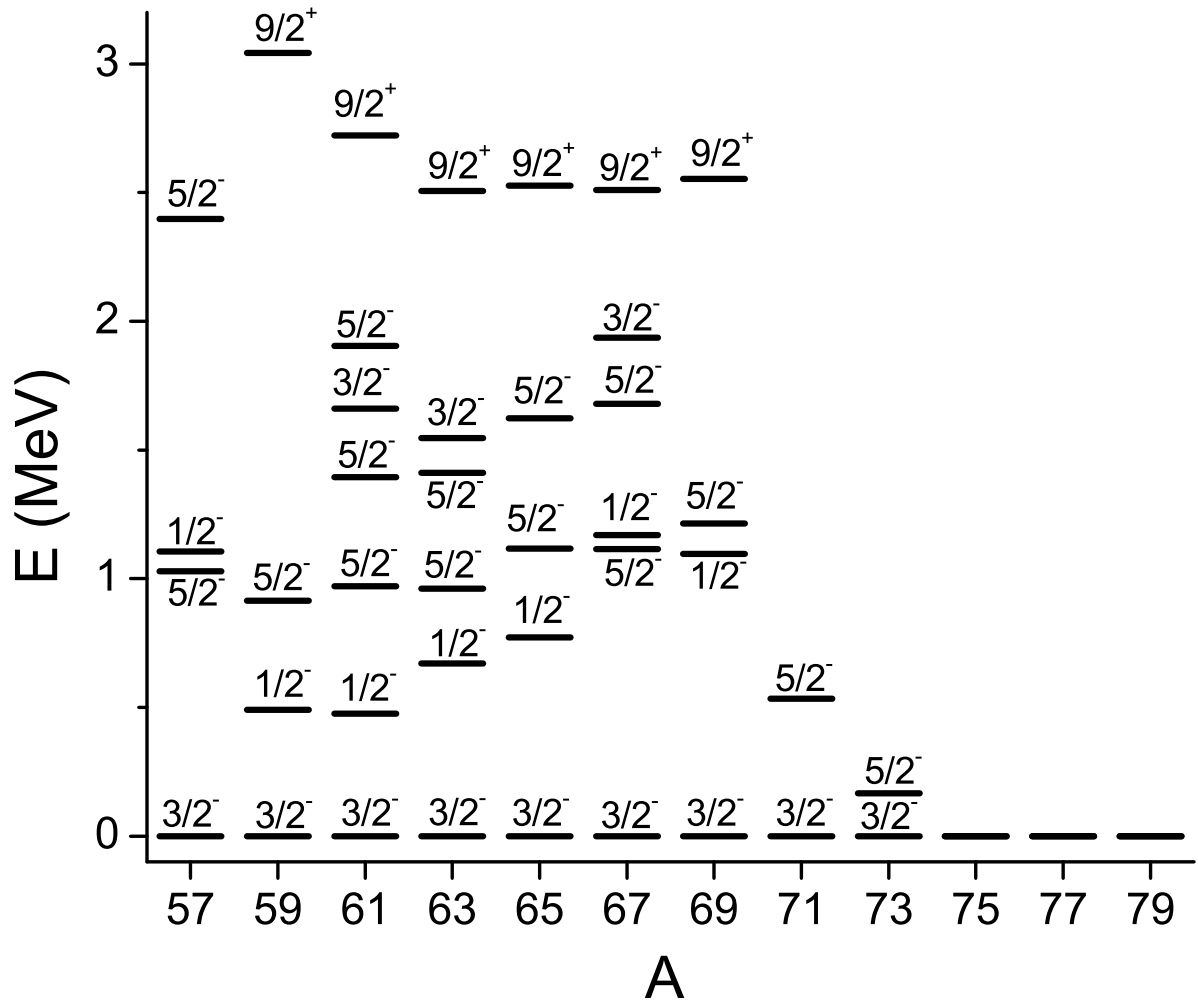}
\caption{Experimental low-lying states in odd-$A$ Cu isotopes~\cite{nndc,Fra01}.}
\end{figure}

\begin{figure}
\includegraphics[width=0.4\textwidth]{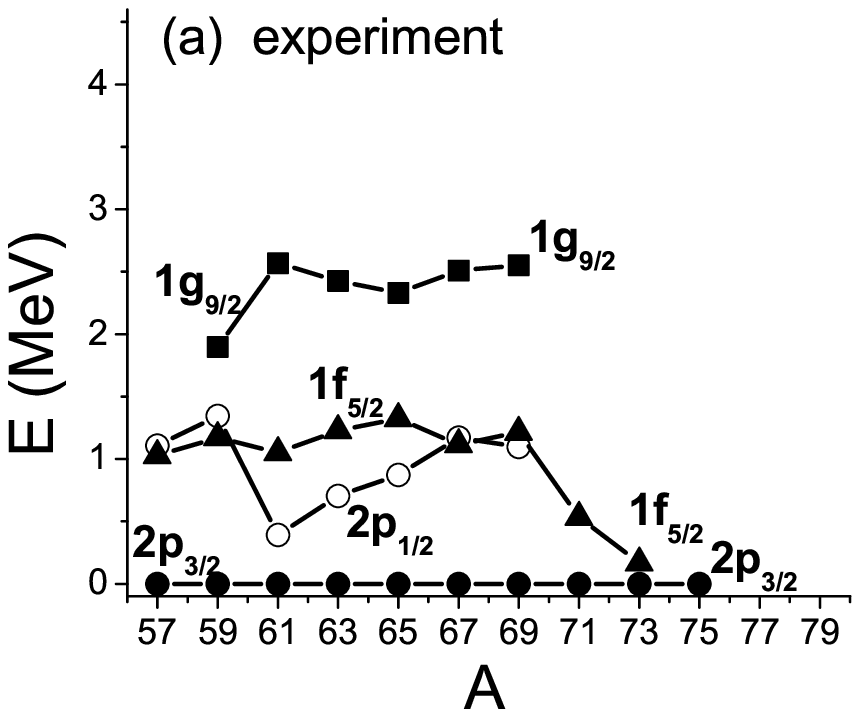}
\includegraphics[width=0.4\textwidth]{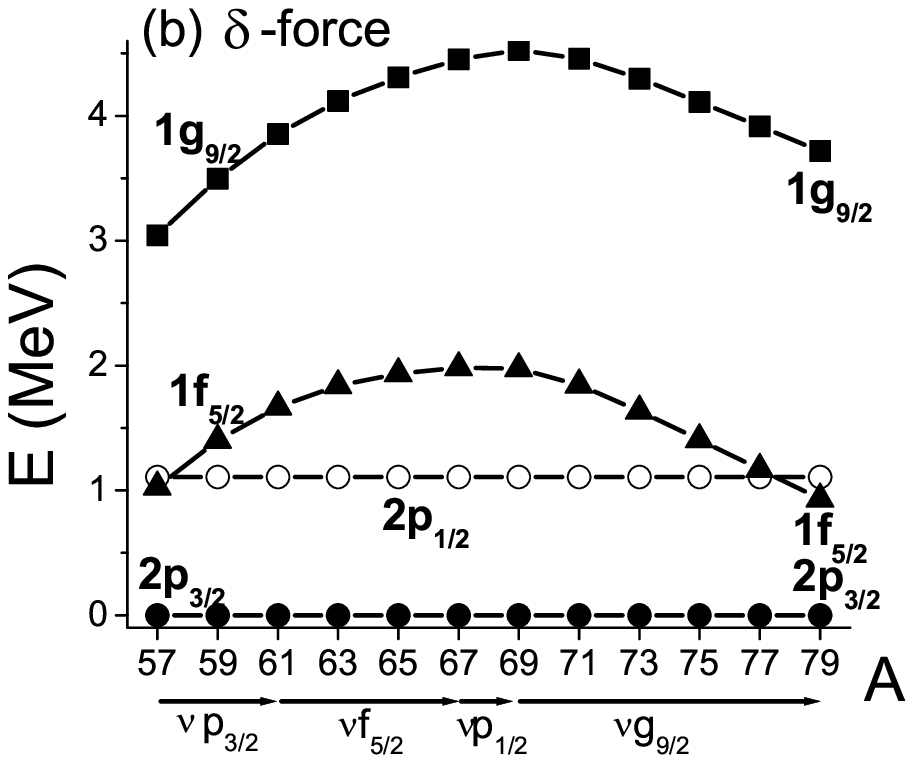}\\
 \hspace{5mm}
\includegraphics[width=0.4\textwidth]{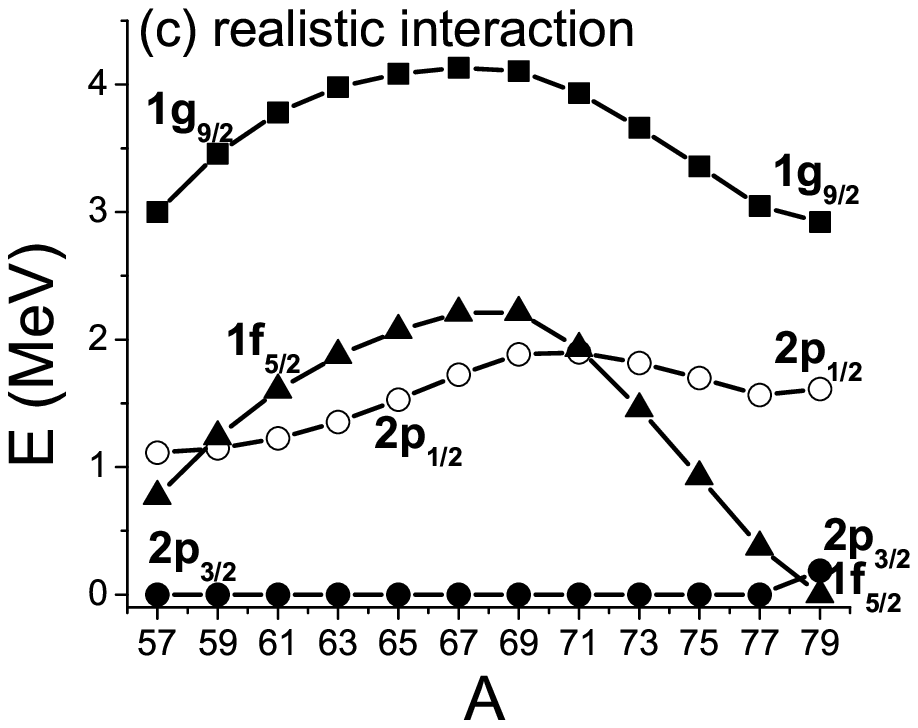}
\includegraphics[width=0.4\textwidth]{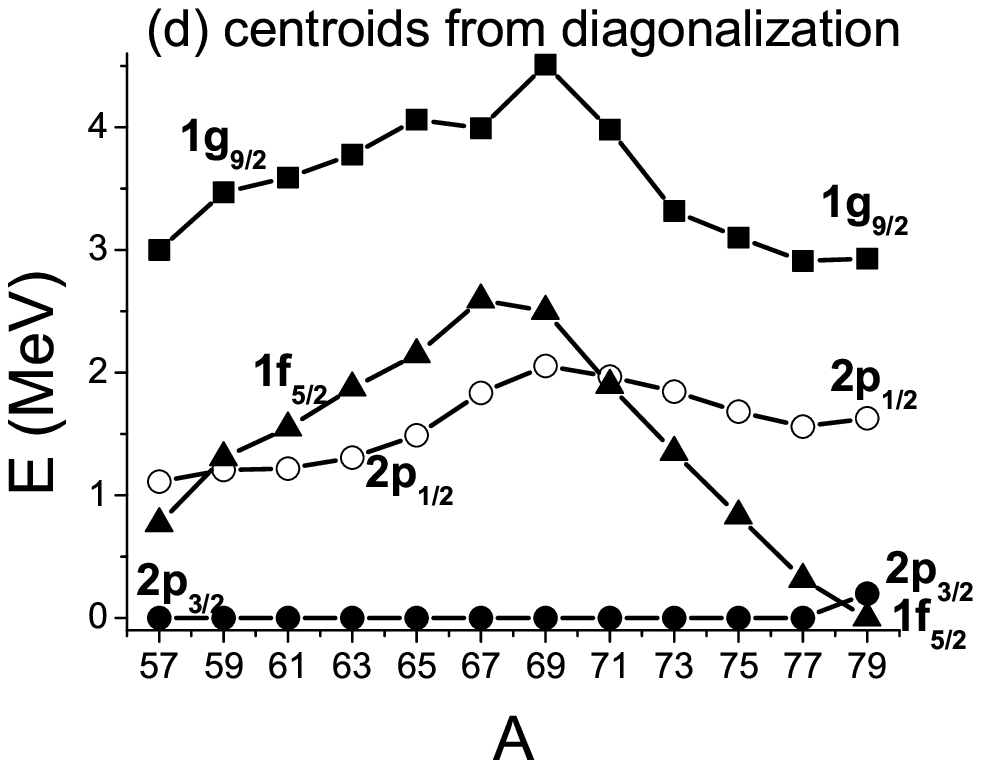}\\ 
\hspace{5mm}
\includegraphics[width=0.4\textwidth]{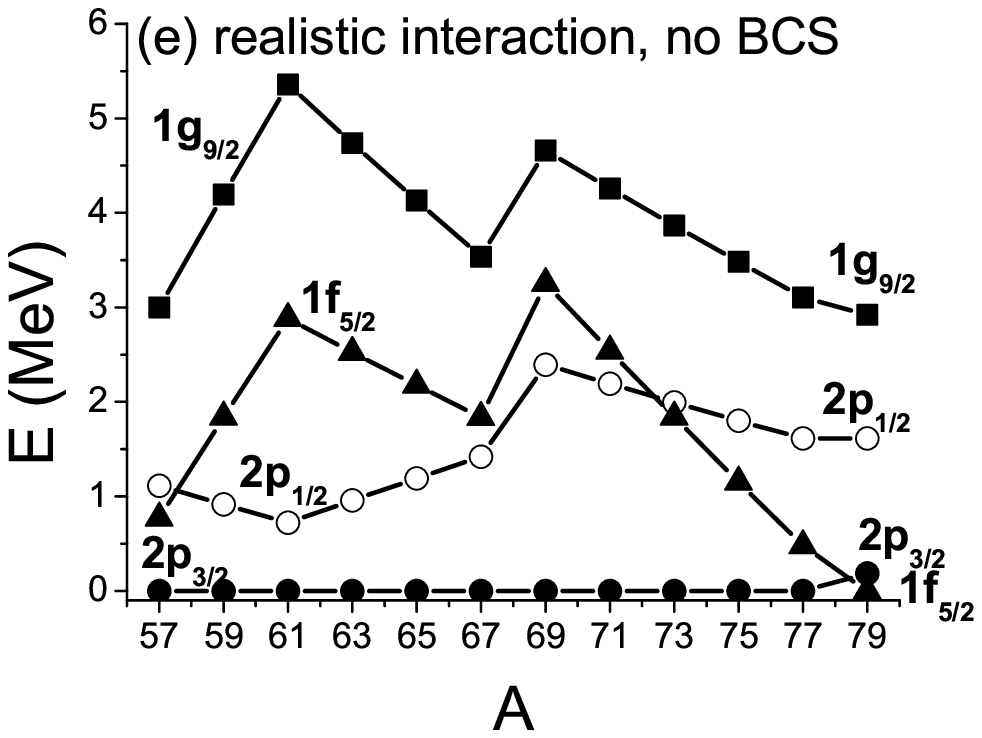}
\includegraphics[width=0.4\textwidth]{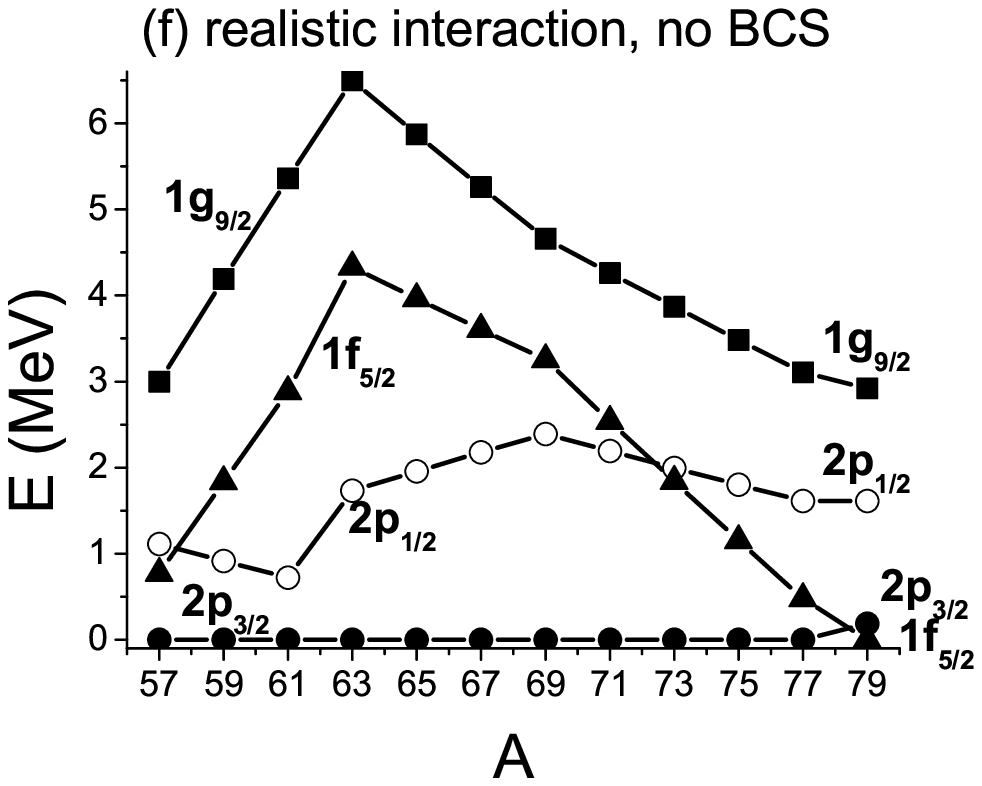}\\
\caption{(a) Experimental centroids when data is available (see Table I) or
lowest states of a given spin in odd-$A$ Cu isotopes;
  (b,c) theoretical evolution of the single-particle states in odd-$A$ Cu
  isotopes due to the monopole part of the proton-neutron $\delta
  $-interaction and realistic interaction, respectively;
  (d) shell-model centroids derived from the exact diagonalization,
  using a realistic interaction; 
  (e,f) the same as in (c), but implying normal filling of the neutron orbitals
  (two possible ways of filling).
  The arrows in figure~(b) below the plot indicate schematically 
  the normal filling of neutron single-particle orbitals in Cu-isotopes.
  See text for details.}
\end{figure}

\begin{figure}
\includegraphics[width=0.4\textwidth]{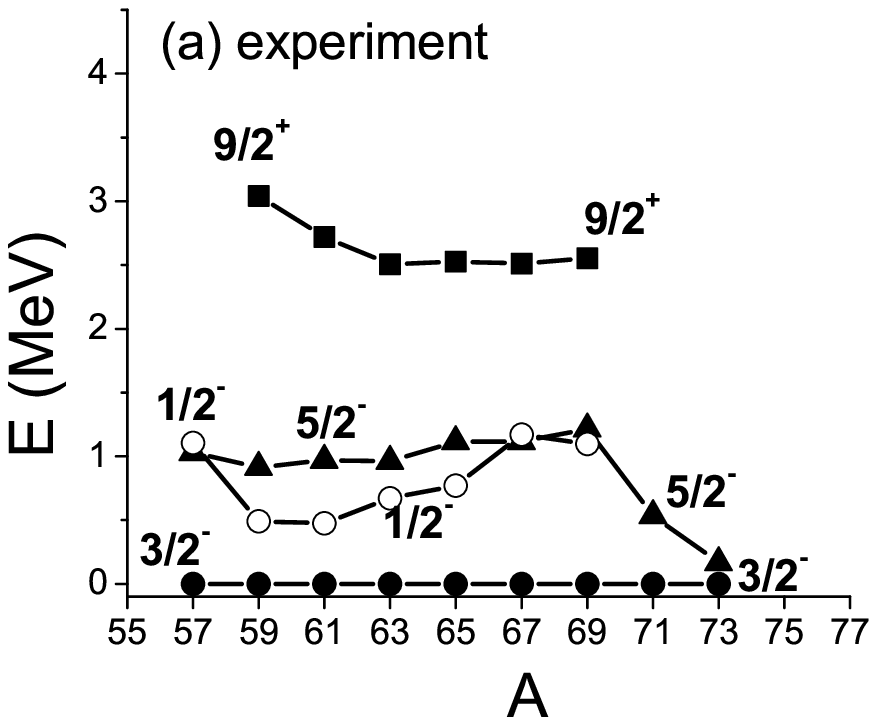}
\includegraphics[width=0.4\textwidth]{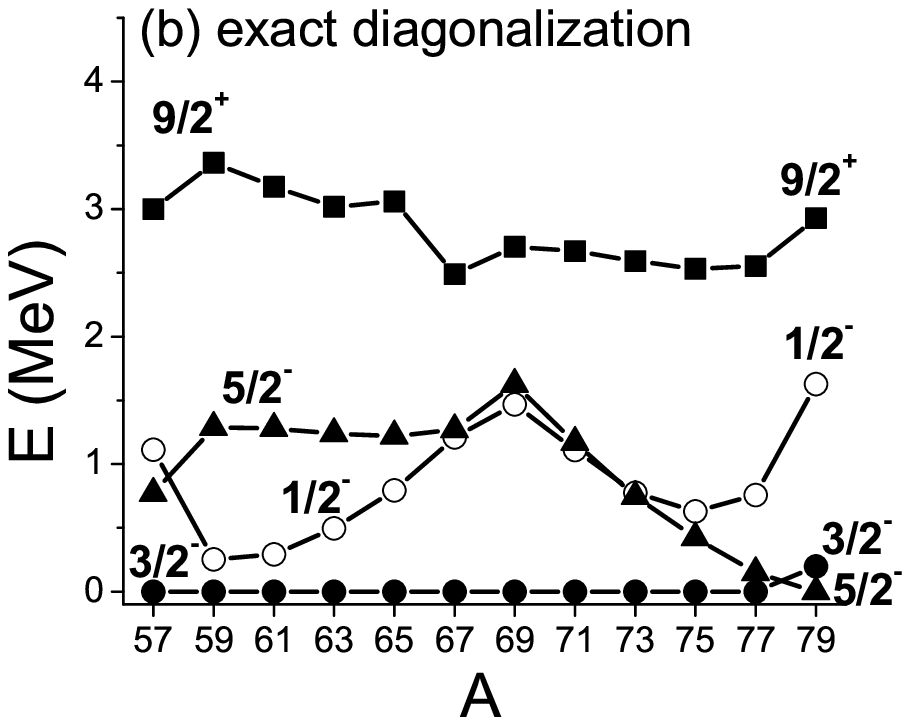}
\caption{The lowest $3/2^-$, $5/2^-$, $1/2^-$ and
  $9/2^+$ states in odd-$A$ Cu isotopes: (a) experimental values;
 (b) as obtained from the shell-model
  diagonalization with the realistic interaction. See text for details.}
\end{figure}

\begin{figure}
\includegraphics[width=0.4\textwidth]{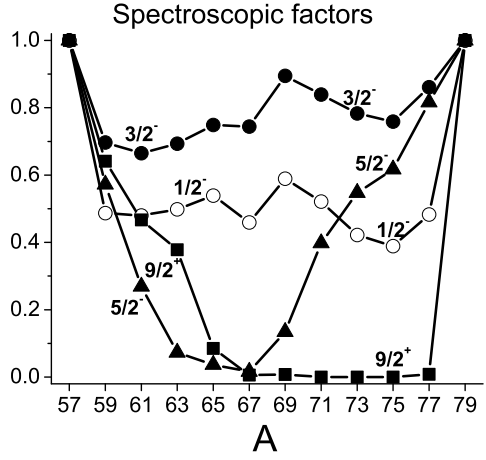}
\caption{Proton stripping spectroscopic factors into the lowest $3/2^-$,
  $5/2^-$, $1/2^-$ and $9/2^+$ states, respectively, in odd-$A$ Cu isotopes
  obtained from the shell-model
  diagonalization with the realistic interaction. The values are normalized to
  the shell-model sum rule. See text for details.}
\end{figure}

\end{document}